\documentclass[lettersize,journal]{IEEEtran}
\usepackage{amsmath,amsfonts}
\usepackage{algorithmic}
\usepackage{algorithm}
\usepackage{array}
\usepackage[caption=false,font=normalsize,labelfont=sf,textfont=sf]{subfig}
\usepackage{textcomp}
\usepackage{stfloats}
\usepackage{url}
\usepackage{verbatim}
\usepackage{graphicx}
\usepackage{cite}
\usepackage{booktabs}
\usepackage{multirow}
\usepackage{bbding}
\newcommand{\coFirstAuthor}{\textsuperscript{*}}
\newcommand{\correspondingAuthor}{\textsuperscript{\dag}}
\begin{document}

\title{CEM-Net: Cross-Emotion Memory Network for Emotional Talking Face Generation}

\author{Kangyi Wu\coFirstAuthor, Pengna Li\coFirstAuthor, Jingwen Fu, Yang Wu, Yuhan Liu, Sanping Zhou,~\IEEEmembership{Member, IEEE}, Jinjun Wang\correspondingAuthor,~\IEEEmembership{Member, IEEE}
\thanks{Kangyi Wu, Pengna Li, Jingwen Fu, Yang Wu, Yuhan Liu, Sanping Zhou, and Jinjun Wang are with the Institute of Artificial Intelligence and Robotics
Xi’an Jiaotong University No.28, West Xianning Road, Xi’an, Shaanxi, China. (email: wukangyi747600@stu.xjtu.edu.cn; sauerfisch@stu.xjtu.edu.cn; fu1371252069@stu.xjtu.edu.cn; wuyang\_cc@stu.xjtu.edu.cn; liuyuhan200095\\@stu.xjtu.edu.cn; spzhou@xjtu.edu.cn; jinjun@mail.xjtu.edu.cn)

This work was supported in part by the National Key Research and Development Project under Grant 2024YFB4708100, National Natural Science Foundation of China under Grants 62088102, U24A20325 and 12326608, and Key Research and Development Plan of Shaanxi Province under Grant 2024PT-ZCK-80.

\coFirstAuthor These authors contribute equally.

\correspondingAuthor Corresponding Author.}}
\markboth{Journal of \LaTeX\ Class Files,~Vol.~14, No.~8, August~2021}%
{Shell \MakeLowercase{\textit{et al.}}: A Sample Article Using IEEEtran.cls for IEEE Journals}

\IEEEpubid{0000--0000/00\$00.00~\copyright~2021 IEEE}

\maketitle

\begin{abstract}
Emotional talking face generation aims to animate a human face in given reference images and generate a talking video that matches the content and emotion of driving audio. However, existing methods neglect that reference images may have a strong emotion that conflicts with the audio emotion, leading to severe emotion inaccuracy and distorted generated results. 
To tackle the issue, we introduce a cross-emotion memory network~(CEM-Net), designed to generate emotional talking faces aligned with the driving audio when reference images exhibit strong emotion. 
Specifically, an Audio Emotion Enhancement module~(AEE) is first devised with the cross-reconstruction training strategy to enhance audio emotion, overcoming the disruption from reference image emotion. Secondly, since reference images cannot provide sufficient facial motion information of the speaker under audio emotion, an Emotion Bridging Memory module~(EBM) is utilized to compensate for the lacked information. It brings in expression displacement from the reference image emotion to the audio emotion and stores it in the memory.
Given a cross-emotion feature as a query, the matching displacement can be retrieved at inference time. Extensive experiments have demonstrated that our CEM-Net can synthesize expressive, natural and lip-synced talking face videos with better emotion accuracy. 
\end{abstract}

\begin{IEEEkeywords}
Audio-Visual, Multimodal, Cross-Emotion Talking Face Generation
\end{IEEEkeywords}

\section{Introduction}
\IEEEPARstart{S}{ynthesizing} realistic talking faces with audio input has gained extensive attention~\cite{yu2020multimodal,chen2023compact,ye2022audio,liu2024multimodal,liu2024osm,zhang2024hierarchical,liu2024audio} and has a wide range of applications, such as digital avatars~\cite{thies2020neural}, video dubbing~\cite{zhang2023dinet} and animation movies~\cite{kim2019neural}. Due to the importance of emotion in human communication~\cite{szajnberg2022face}, an increasing number of researchers have begun to focus on generating emotion-controllable talking faces~\cite{goyal2023emotionally} in recent years. Generally, emotional talking face generation is driven by three inputs: speaker images, audio input, and target emotion.

\begin{figure}[t]
    \centering
\includegraphics[width=\linewidth]{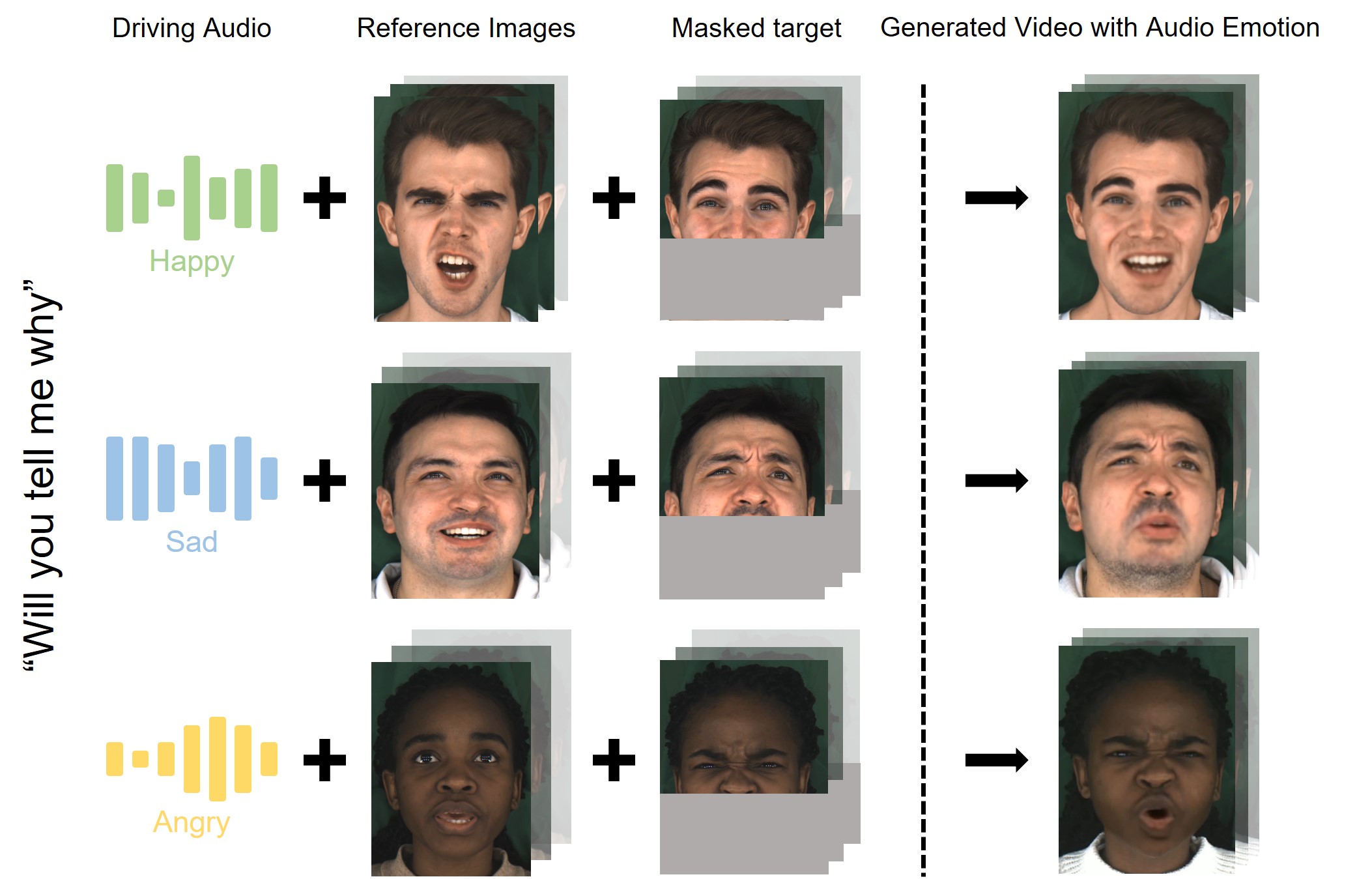}
    \caption{Audio-driven talking face video generated by the proposed CEM-Net. Given an emotional audio clip, CEM-Net utilizes appearance prior of reference images to synthesise the lower-half face. Even if the reference has apparent emotion prior, CEM-Net is capable of generating a realistic talking face video, where lip motions and emotion are coherent with the driving audio.}
\label{fig:illustration}
\end{figure}

\begin{figure}[!t]
  \centering
\includegraphics[width=\linewidth]{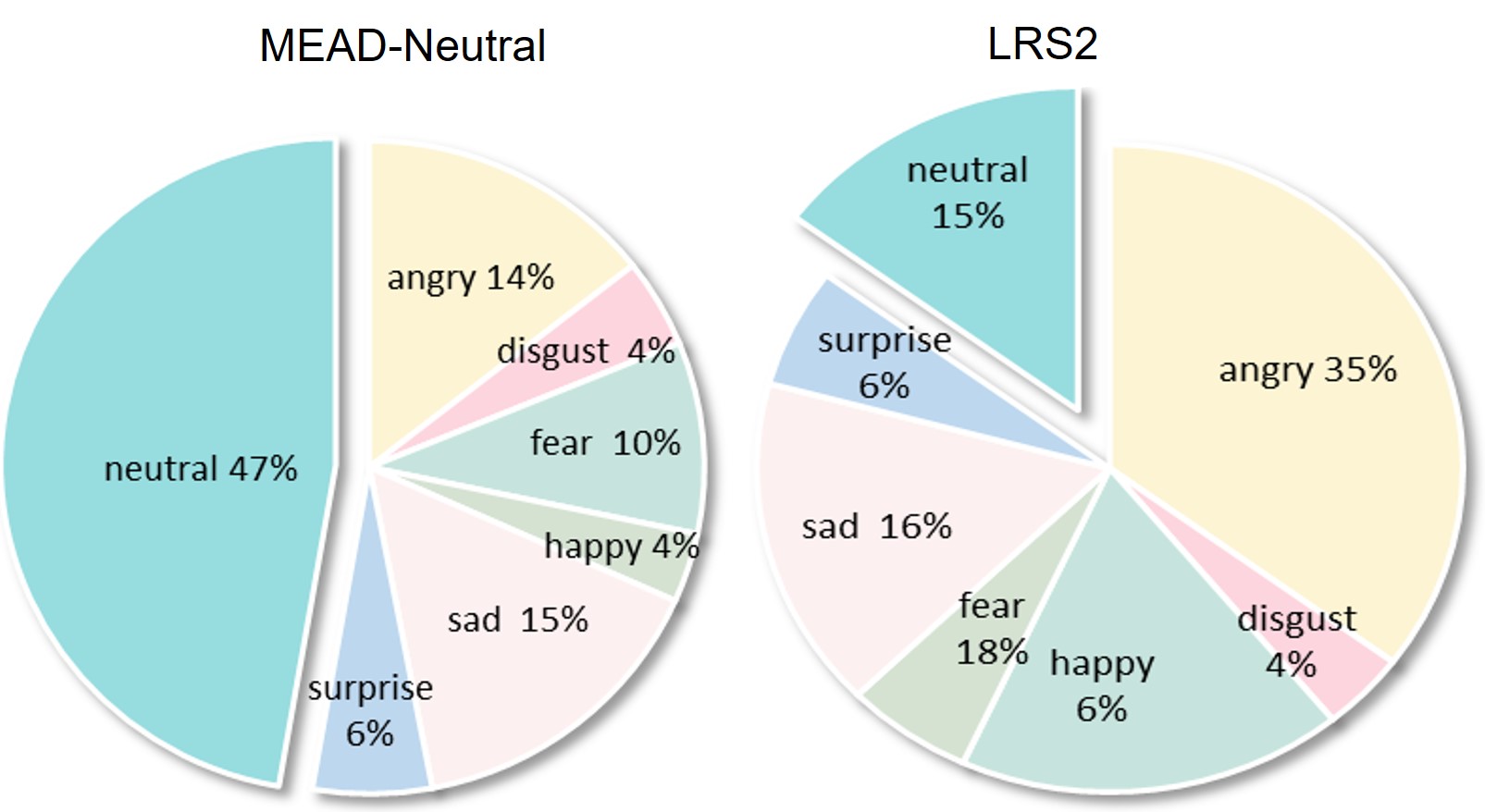}
  \caption{Emotion proportion for MEAD \textbf{Neutral} videos and all videos in the LRS2 dataset.}
  \label{fig:emo_proportion}
\end{figure}

Existing methods primarily derive the target emotion from an arbitrary emotion label\cite{xu2023multimodal, feng2024emospeaker,gururani2023space,sheng2023stochastic}, additional emotional videos~\cite{agarwal2023audio} and emotional audio\cite{ji2021audio,tan2023emmn}. Among them, the first two sources share the same problem: it's difficult to choose the suitable target emotion to make the generated result semantically and visually realistic~\cite{sun2024avi,ji2021audio}. 
Differently, audio inherently contains the speaker's emotion information.
So in this paper, we are dedicated to obtaining the target emotion from audio.
As depicted in Fig.~\ref{fig:illustration}, our goal is to animate a speaker's face in still reference images and generate a talking video coherent with the content and emotion of driving audio. Previous methods commonly use audio emotion to refine faces predicted from driving audio and reference images\cite{eskimez2021speech,liang2022expressive,ma2023talkclip}.
However, in real-world deployment, reference images may have a strong emotion that conflicts with audio emotion. This will result in distortion and incorrect expression of generated talking videos. 
Therefore, only when purely neutral faces are provided as references can the result be synthesized with accurate emotion. However, it's difficult to acquire purely neutral faces for the two following reasons:

\IEEEpubidadjcol\textbf{$(i)$~Purely neutral faces only constitute a small portion in reality.} As illustrated in Fig.~\ref{fig:emo_proportion}, proportion of seven emotions are calculated with Deepface\cite{serengil2021lightface} for \textbf{neutral videos} in the MEAD dataset\cite{wang2020mead} and \textbf{all videos} in the LRS2 dataset\cite{son2017lip}. Even in the neutral videos from the MEAD dataset, frames with neutral emotion only account for 47.25\%. For the LRS2 dataset which is collected from BBC videos and is closer to real-world scenarios, the ratio for neutral emotion is only 14.82\%. This illustrates that people tend to express emotion when they are talking, and reference images, to varying degrees, exhibit some emotional bias towards an arbitrary emotion. It's possible that we can require the user to only upload neutral references. However, according to~\cite{gabriel1990worse}, in software engineering, ``Simplicity is the most important consideration." It's better to allow users to generate realistic talking faces in natural way. 

\IEEEpubidadjcol \textbf{$(ii)$~Acquiring neutral faces with face emotion editing models~\cite{shen2020interpreting,azari2024emostyle,nitzan2022mystyle} is costly and the generated results are not necessarily good.} We evaluate the loss of identity information with the Identity Deterioration metric of edited face images by Arcface\cite{deng2019arcface}. The closer this number is to 0, the more identity information is preserved. As can be seen from Table~\ref{tab:emo_editing}, face editing models typically encounter identity information decay which cannot be compensated for in the talking face generation stage. \textit{e.g.} For EmoStyle\cite{azari2024emostyle}, the identity information loss of the edited neutral face is 0.12. If talking faces are generated on the edited neutral face, the final identity deterioration will be definitely higher than 0.12, which is unacceptable and will lead to severe artifacts. Moreover, the computational load of face emotion editing models is significant, some even exceeding that of our talking face generation model.

Based on the observation above, how to achieve emotional talking face generation when reference images have a strong emotion conflicting with that from the driving audio, which is called ``cross-emotion", is an urgent problem to be solved. There exist two unavoidable challenges when neutral faces are not available:  1) Since our target emotion is extracted from driving audio, it can be easily disturbed by the timbre of different speakers. For example, the difference between an angry audio and a happy audio may be smaller than that between two different speakers with the same emotion. It's crucial to enhance the emotional information from the voice. 2) In a cross-emotion setting, due to the great differences between emotions, reference images fail to provide sufficient facial motion information to generate talking faces under the target emotion. Also, different speakers have variant motion habits. Models are required to learn the unique motion habits of a specific speaker under a certain emotion. 

\begin{table}
\centering
  \caption{The decay of identity information after face emotion editing, coined as Identity Deterioration, and model parameter of three popular face emotion editing models.}  
\renewcommand\arraystretch{1.5}
    \tabcolsep=0.5cm
  \begin{tabular}{c|c|c}
    \toprule
    Method&\textbf{Identity Deterioration}$\downarrow$ &Parameter\\
    \hline
    Interface\cite{siarohin2019animating}&0.19&23.08M\\
    MyStyle\cite{nitzan2022mystyle}&0.21&28.26M\\
    EmoStyle\cite{azari2024emostyle}&\textbf{0.12}&48.40M\\
    \bottomrule
  \end{tabular}

  \label{tab:emo_editing}
\end{table}
In this paper, we introduce a novel audio-driven cross-emotion talking face generation framework, called Cross-Emotion Memory Network~(CEM-Net). \textbf{To cope with the first challenge}, we carefully design an Audio Emotion Enhancement module~(AEE) that decouples audio signal into timbre, content and emotion. Specifically, cross-reconstruct training strategy is adopted to train this module inspired by ~\cite{zhang2022self}. After the AEE module is trained, a stronger target emotion is acquired. \textbf{To address the second challenge}, we resort to external memory network\cite{weston2014memory} to compensate for the lacked motion information and speakers' motion habits. In detail, the Emotion Bridging Memory Network~(EBM) is implemented to map the cross-emotion features to expression displacement. The memory first stores the representative expression displacements. By leveraging the various combinations of cross-emotion features, we can obtain different expression displacements for different motion habits. In this way, the lacked information can be compensated for by memory. 

Our contributions are summarized as follows:
\begin{itemize}
\item For the first time, we indicate that the reference image in reality usually contains a certain amount of emotional information, which, if conflicts with the target emotion, will eventually make the mouth shape of the generated results emotionally inaccurate and distorted.  

\item A brand-new framework, CEM-Net, is introduced for the ``cross-emotion" talking face generation task. We devised an Emotion Bridging Memory Network~(EBM) to compensate for the lacked motion information under target emotion and speakers' motion habits in reference images. Also, an audio emotion enhancement module~(AEE) is utilized to strengthen the audio emotion. 
\item We systematically evaluate the emotion correctness and lip synchronization of the results and extensive experiments demonstrated the superiority of our method over state-of-the-art methods. 
\end{itemize}

\section{Related Work}
\subsection{Audio-driven talking face generation.} Audio-driven talking face generation is to synthesize a sequence of video frames of a certain identity whose lip movement is synchronized with the audio input. These techniques are broadly categorized into two types: person-specific and person-generic methods. Although person-specific approaches deliver superior animation outcomes, their application scenarios are restricted. Ad-Nerf~\cite{guo2021ad} generates not only the head region but also the upper body via two individual neural radiance fields. Facial~\cite{zhang2021facial} proposes a FACIAl-GAN to synthesize 3D face animation with realistic motions of lips, head poses and eye blinks. However, these approaches all necessitate videos of the target speaker for re-training or fine-tuning, a requirement that might be unattainable in real-world situations. Thus developing person-generic methods capable of synthesizing talking face videos for speakers not previously seen is of greater importance. MCNET~\cite{hong2023implicit} learns a global facial representation space and design an implicit identity representation conditioned memory compensation network for talking head generation. IP-LAP~\cite{zhong2023identity} devises a transformer-based landmark generator and a new alignment module to enhance identity information. MODA~\cite{liu2023moda} proposes a dual-attention module to learn the correlation between lip-sync and other movements. Though these methods can generate videos with high fidelity, they fail to account for the emotional factor, which is crucial for achieving realism in the generated videos and ensuring the comprehensive conveyance of semantic meaning from the driving audio.

\subsection{Emotional talking face generation.} Since emotional talking face generation methods can synthesize more expressive results, it has attracted considerable attention in recent years. Based on the emotion source, emotional talking face generation methods can be generally categorized as vision-driving or audio-driving types. Vision-driving methods extract emotion representation from emotional videos or images. EAMM~\cite{ji2022eamm} proposes a two-stage framework and devises an Implicit Emotion Displacement Learner to add emotion displacement. PDFGC~\cite{wang2023progressive} devises a progressive disentangled representation learning strategy to realize fine-grained control over multiple perspectives. However, selecting appropriate driving source to ensure both semantic and visual realism in the results is labor-intensive in practice. It's more significant to generate emotional talking face videos whose emotion is directly predicted from the driving audio. EVP~\cite{ji2021audio} applies a cross-reconstruction emotion disentanglement module to extract emotion from audio. SadTalker~\cite{zhang2023sadtalker} presents an ExpNet to learn accurate facial expressions from audio by distilling both coefficients and 3D-rendered faces. However, existing methods all add this emotion knowledge to intermediate representation predicted from driving audio and reference images with neutral faces, which is difficult to acquire in most cases. If reference identity images have strong emotion different from that in driving audio, there will be severe emotional inconsistency between generated videos and driving audio. On the contrary, we aim to generate talking faces videos with audio emotion when reference images have another disruptive emotion.

\subsection{Memory Network for talking face generation.} Memory Network~\cite{weston2014memory} utilizes inference components in conjunction with a long-term memory component for reasoning. Due to its versatile capacity to store, abstract, and organize long-term knowledge into a coherent form, memory network is favoured in several tasks~\cite{yoo2019coloring,huang2021memory,sun2021mamba,fei2021memory}, including talking face generation. EMMN~\cite{tan2023emmn} stores emotion embedding and lip motion in memory, together with paired expression code to make lip movement consistent with that of expression. MCNet~\cite{hong2023implicit} learns a unified spatial facial meta-memory bank to provide rich facial structure and appearance priors. SyncTalkFace~\cite{park2022synctalkface} proposes an audio-lip memory to compensate for visual information corresponding to input audio and enhance fine-grained audio-visual consistency. However, previous works all focus on emotion-agnostic or emotional talking face tasks with neutral faces as references and fail to adapt effectively to cross-emotion generation. For the memory network, compensating for the lacked motion information of reference images has yet to be attempted. We build a carefully designed emotion-bridging memory module to store and align cross-emotion features and expression displacement, which is utilized to compensate for the motion information and speakers’ motion habits. In this way, our memory network will provide warping information between two arbitrary emotions in consideration of the identity information.

\section{Method}
\begin{figure*}[t]
    \centering
    \includegraphics[width=\linewidth]{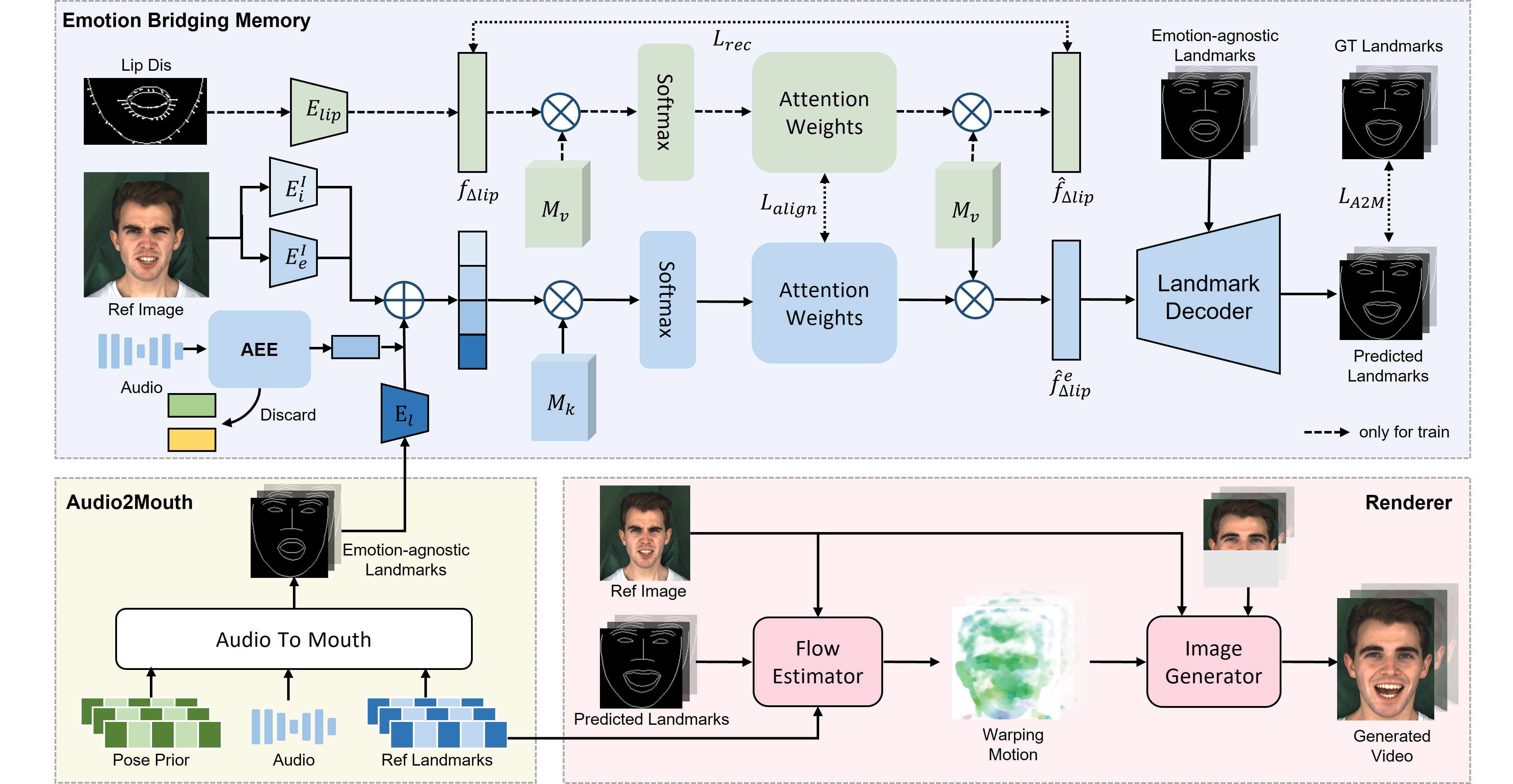}
    \caption{Overview of the proposed method. The Audio2Mouth module predicts lip landmarks from reference images and driving audio without considering the emotion. Then the Emotion Bridging Memory module produces the cross emotion lip displacement based on the emotion embedding from reference images and driving audio. Lastly, the Renderer generates lip-synced and high fidelity results. The structure of AEE is presented in Fig.~\ref{fig:IED}.}
    \label{fig:framework}
\end{figure*}
The framework of our proposed CEM-Net is depicted in Fig.~\ref{fig:framework}. The first is Audio2Mouth module which predicts emotion-agnostic lip landmarks from reference images and driving audio. Then the target emotion is disentangled from driving audio with the Audio Emotion Enhancement module. Thirdly, an image emotion extractor trained with data augmentation strategies~\cite{ji2022eamm} extracts source emotion from the reference image. Next, the identity feature of the reference image extracted from ArcFace~\cite{deng2019arcface} pre-trained on Casia ~\cite{yi2014learning} is used as an additional identity condition. The three above combined with the emotion-agnostic lip motion form the final cross-emotion features, which drive the Emotion Bridging Memory. Given the cross-emotion feature as a query, we can obtain the corresponding emotional lip displacement to add emotional movement. Lastly, the renderer is incorporated to generate lip-synced and high-fidelity results with identity information well-preserved. Each part of our method is illustrated in detail in the following sections.

\subsection{Audio2Mouth.}
An audio2mouth module is firstly built to map reference images, audio content and pose landmarks to landmarks without considering emotion factor. Since transformer~\cite{vaswani2017attention} has demonstrated its superiority in learning long-term relations, we choose the transformer encoder as the backbone. $N$ randomly selected reference identity images are used for predicting $T$ adjacent frames at a time. Specifically, for each reference identity image $\{I_i^r\}^N$, landmark detector $L$ is used to predict the initial motion representation $\{l_i^r\}^N$ first. Then, three encoders are constructed to encode reference landmarks $\{l_i^r\}^N$, driving audio $\{a_t\}_{t=1}^T$ and pose landmarks $\{p_t\}_{t=1}^T$ respectively, generating reference embedding $\{e_i^r\}^N$, audio embedding $\{e^a_t\}_{t=1}^T$ and pose embedding $\{e^p_t\}_{t=1}^T$. 
All the features serve as the input to Audio2Mouth model $A2M$. The last T output tokens are adopted to predict mouth motions.
Generally, the function of the Audio2Mouth module can be formulated as:
\begin{equation}
    \{z_k\}_{k=1}^{N+2T} = A2M(\{e_i\}^{N+T+T}),
\end{equation}
\begin{equation}
    \overline{m}_t = Linear(z_{N+T+t}) \quad t=1,2,3,\ldots,T.
\end{equation}
To update the Audio2Mouth module, we minimize the $L_1$ distance between the predicted $\overline{m}_t$ and ground truth $m_t$. Also, to enhance the smoothness over time, we implement a continuity regularization technique to ensure the consistency of predicted mouth movements between $\overline{m}_{t+1} - \overline{m}_{t}$ and $m_{t+1} - m_t$, which can be formulated as:
\begin{equation}
    L_v = \frac{1}{T-1}\sum\nolimits_{t=1}^{T-1}\left \vert \left \vert (\overline{m}_{t+1}-\overline{m}_{t}) - (m_{t+1}-m_{t})\right\vert\right\vert_2.
\end{equation}
Therefore, the loss function for the Audio2Mouth module is derived by:
\begin{equation}
    L_{A2M}=L_1 + \lambda L_v,
\end{equation}  
where $\lambda$ serves as a hyper-parameter to do a trade-off between precision and smoothness.

\begin{figure}[!t]
  \centering
\includegraphics[width=\linewidth]{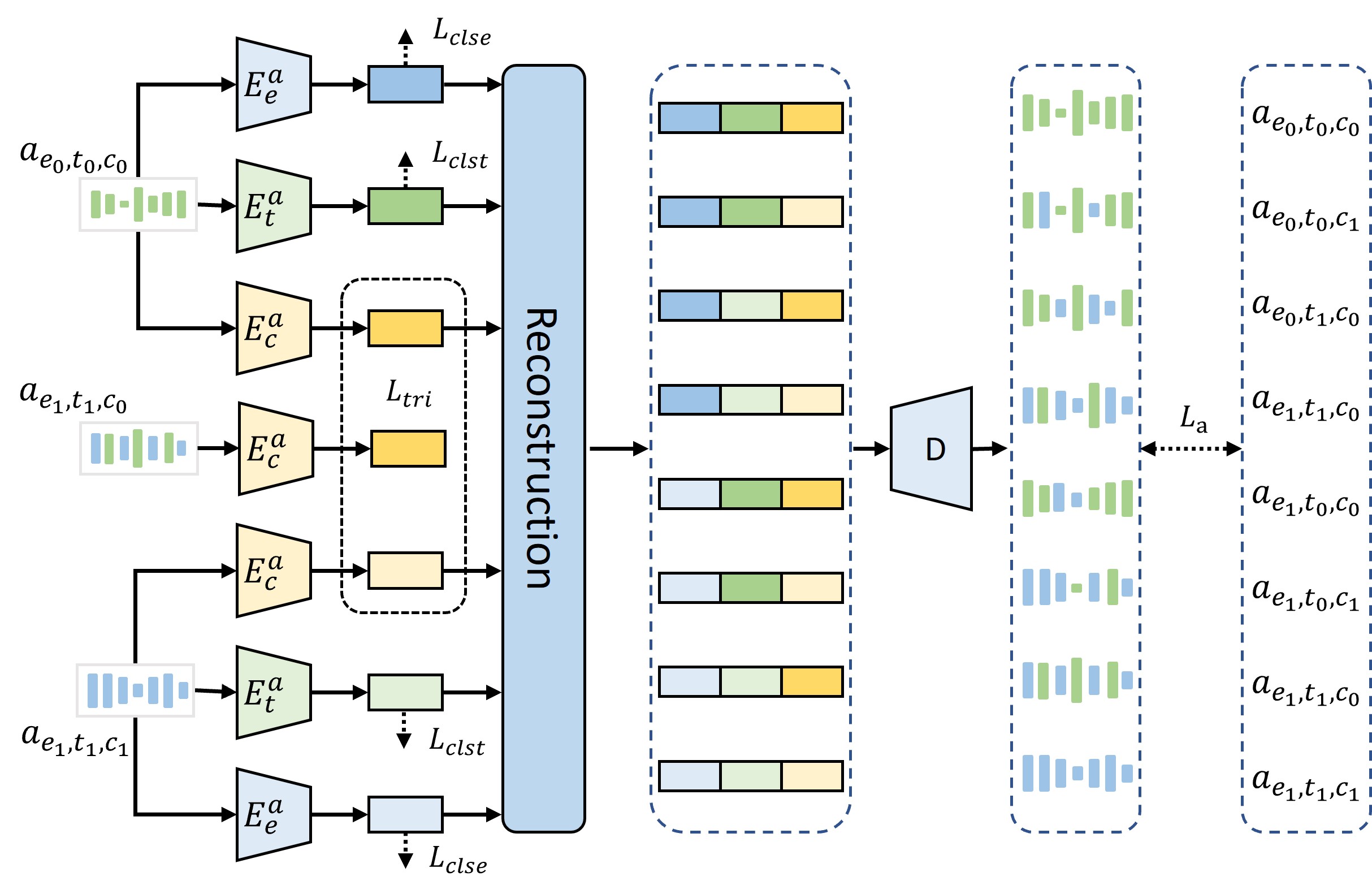}
  \caption{Audio Emotion Enhancement~(AEE) with cross-reconstruction training strategies. We extract disentangled different timbre, content and emotion from two individual audios and reconstruct new audios.}
  \label{fig:IED}
\end{figure}

\subsection{Audio Emotion Enhancement.} 
Accurately extracting emotional information from the driving audio is important for cross-emotion talking face generation. Although previous works~\cite{ji2021audio,tan2023emmn} have made their efforts and achieved great face emotion recognition scores, they have difficulty dealing with person-generic tasks due to speaker variations. Instead of simply disentangling emotion from audio leading to emotion knowledge degraded by the pronunciation habits of different individuals, we construct a brand-new Audio Emotion Enhancement~(AEE) module with cross-reconstruction~\cite{aberman2019learning} technique to enhance audio emotion to eliminate the influence of timbre~(identity) and content from audio.
To implement cross-reconstruction training, we achieve paired audio data with the different emotions and content spoken by individual speakers from MEAD~\cite{wang2020mead}. The details of data processing are put in the Appendix. A temporal alignment algorithm\cite{berndt1994using} is utilized to synchronize speeches of varying lengths.
As shown in Fig.~\ref{fig:IED}, three encoders $E^a_e$, $E^a_t$ and $E^a_c$ are leveraged as emotion encoder, timbre encoder and content encoder to extract disentangled emotion embedding, timbre embedding and content embedding and from a specific audio clip $a_{e_i,t_j,c_k}$ with emotion $i$, identity $j$ and content $k$. Specifically, two audio clips, $a_{e_i,t_j,c_k}$ and $a_{e_m,t_p,c_q}$, serve as an input sample. We concatenate $E_t (a_{e_i,t_j,c_k})$, $E_e (a_{e_m,t_p,c_q})$ and $E_c (a_{e_i,t_j,c_k})$ together to form a new audio embedding and reconstruct the audio clip $a_{a_{e_i,t_p,c_k}}$ with a decoder $D$. 
To make the three encoders have disentanglement capability and avoid loss of knowledge, we update the model with four losses: emotion classification loss, identity classification loss, content reconstruction loss, and audio reconstruction loss. 
Given $a_{e_0,t_0,c_0}$ and $a_{e_1,t_1,c_1}$ as input, the audio reconstruction loss is defined by:
\begin{multline}
L_{a}=\sum_{i,j,k=0}^1 \Bigl\vert \Bigl\vert D(E_t(a_{e_i,t_i,c_i}),E_e (a_{e_j,t_j,c_j}),\\E_c (a_{e_k,t_k,c_k}))-a_{e_i,t_j,c_k} \Bigr\vert \Bigr\vert_2.
\end{multline}

With additional classifiers $C_t$ and $C_e$, identity classification loss $L_{clst}$ and emotion classification loss $L_{clse}$ are adopted to map audio with the same emotion category or same identity into clustered groups in latent space. 
What's more, triplet loss is used to make audio with the same content share similar content embedding:
\begin{multline}
    L_{tri} = \sum_{i,j,k=0}^1 max(\alpha+\left\vert \left\vert E_c(a_{e_i,t_i,c_i}-E_c(a_{e_k,t_k,c_k})\right\vert \right\vert \\
    -\left\vert \left\vert E_c(a_{e_i,t_i,c_i}-E_c(a_{e_{1-k},t_{1-k},c_{1-k}})\right\vert \right\vert,0).
\end{multline}

The summarization of the losses above is the final objective of the AEE module:
\begin{equation}
    L_{AEE} = \lambda_{a} L_{a} + \lambda_{clst} L_{clst} + \lambda_{clse} L_{clse} + \lambda_{tri} L_{tri},
\end{equation}
where $\lambda_s$ are hyper-parameters for balancing the different weights of losses. The AEE loss enhances the disentanglement among emotion, timbre and content features. After the AEE module training is finished, the trained audio emotion encoder $E^a_e$ is adopted to extract clean target emotion unaffected by timbre and content from audio.

\subsection{Emotion Bridging Memory.} 
To further deal with emotional conflict, we combine target emotion with source emotion extracted from reference images to form cross-emotion features. Besides, different speakers have specific motion habits so the identity information should be considered. We bring in identity features from reference images to cross-emotion features. Due to the great gap between the source and target emotion, reference images fail to provide sufficient motion information to generate talking faces, we propose the Emotion Bridging Memory Network~(EBM) to compensate for the expression motion information and speakers' motion habits. Specifically, EBM stores and aligns the cross-emotion and lip displacement features, where lip displacement is obtained from the difference between ground-truth and emotion-agnostic landmark. EMB module consists of key memory~$\mathcal{M}_{k}\in{\mathcal{R}^{K\times{D}}}$ and value memory~$\mathcal{M}_{v}\in{\mathcal{R}^{K\times{D}}}$, where K is memory size and D is dimension of each slot. In detail, the value memory learns to store representative lip displacement features. We firstly obtain lip displacement features~$f_{\Delta{lip}}\in{\mathcal{R}^D}$ using a lip landmark encoder. Then attention weights are computed by softmax of cosine similarity between~$f_{\Delta{lip}}$ and each slot as follows:
\begin{equation}
    {\alpha}_{v} = {Softmax}({f_{\Delta{lip}}}\cdot{\mathcal{M}_{v}^t}).
\end{equation}

Through the attention weights, we reconstruct the lip displacement features by:
\begin{equation}
    \hat{f}_{\Delta{lip}} = \sum\nolimits_{i=1}^K{\alpha}_{v}^i\cdot{m_{v}^i} = {\alpha}_{v}{\mathcal{M}_{v}},
\end{equation}
where $m_{v}^i$ is the $i$th slot in $\mathcal{M}_{v}$. The reconstruction loss to optimize~$\mathcal{M}_v$ is formulated as:
\begin{equation}
    L_{rec} = {\left\vert \left\vert f_{\Delta{lip}}-\hat{f}_{\Delta{lip}}\right\vert\right\vert}^2.
\end{equation}

By this means, typical lip displacement features can be stored in $M_v$. However, at inference time, only the cross-emotion features are available. Given a cross-emotion feature $f_e$ as a key, we expect the obtained value to be the corresponding lip displacement feature. In other words, we should ensure that attention weights of key memory and value memory are as close as possible. Therefore, we adopt KL divergence to align the attention weights:
\begin{equation}
    {\alpha}_{k} = {Softmax}({f_e}\cdot{\mathcal{M}_{k}^t}),
\end{equation}
\begin{equation}
    L_{align} = KL({\alpha}_{k}\vert\vert{\alpha}_{v}).
\end{equation}

Aligning key attention weights and value attention weights, we obtain corresponding lip displacement features by:
\begin{equation}
    \hat{f}_{\Delta{lip}}^e = \sum\nolimits_{i=1}^K{\alpha}_{k}^i\cdot{m_{v}^i} = {\alpha}_{k}{\mathcal{M}_{v}}.
\end{equation}

Then, we concatenate $\hat{f}_{\Delta{lip}}^e$ and emotion-unaware lip motion obtained by our Audio2Mouth to generate final emotion-aware mouth motions via a landmark decoder and utilize $L_{A2M}$ loss to supervised the training of this module.

\subsection{Renderer.} 
Inspired by Monkey-Net~\cite{siarohin2019animating}, our renderer consists of a flow estimator and an image generator. Given the predicted landmarks from EBM, the flow estimator takes the original reference images and their landmarks as input to predict the warping motion first. Then a generator is constructed to synthesize the final result based on the warping motion. Specifically, we concatenate refined mouth landmarks $\{\hat{m}_t\}_{t=1}^T$ and pose landmarks $\{p_t\}_{t=1}^T$ to form predicted face landmarks $\{\hat{f}_t\}_{t=1}^T$. They will be fed to flow estimation module $W$ together with reference landmarks $\{l_i^r\}^N$. and reference images $\{I_i^r\}^N$. For each reference landmark $l_i^r$, T motion fields $M^i_{1\rightarrow T}$are generated. The formulation of the flow estimation module is:
\begin{equation}
    M^i_{1\rightarrow T} = W(l_i^r, \hat{f}_{1\rightarrow T}, I_i^r) \quad i=1,2,3,\ldots,N.
\end{equation}

The warped reference images for each predicted frame are calculated as:
\begin{equation}
    \overline {I_t^r} = \frac{\sum_{i=1}^N \omega_t^i M^i_{t}(I_i^r)}{\sum_{i=1}^N \omega_i}\quad t=1,2,3,\ldots,T,
\end{equation}
where $\omega_t^i$ represent the predicted weight for $I_i^r$ when generating frame $t$. 
At last, the warped reference images $\{\overline {I_t^r}\}_{t=1}^T$, predicted face landmarks $\{\hat{f}_t\}_{t=1}^T$ and the masked target face $I_t^m$ are concatenated as input to the generator $G$. The overall process of the generator is formulated as:
\begin{equation}
    \hat{I}_t = G(\overline {I_t^r}, I_t^m, \hat{f}_t)\quad t=1,2,3,\ldots, T.
\end{equation}

The renderer is trained in an end-to-end manner, with perceptual loss, GAN loss and feature matching loss implemented. Perceptual loss $L_{per}$~\cite{johnson2016perceptual} is employed to ensure the precision of the motion field and the quality of the generated faces. 
GAN loss $L_{gan}$ and feature matching loss $L_{fm}$ from ~\cite{wang2018high} is also utilized to improve image realism.
In total, the loss of renderer is formulated as:
\begin{equation}
    L_{renderer} = \omega_{per}L_{per} + \omega_{gan}L_{gan} + \omega_{fm}L_{fm}.
\end{equation}

\section{Experiments}
\begin{table*}[!t]
    \centering
        \caption{Quatitative comparison with state-of-the-art audio-driven talking face generation methods on neutral faces based and cross emotion settings on the MEAD dataset. In neutral base settings, neutral faces are provided as reference images. But in cross emotion settings, reference images have randomly chosen emotions different from those of driving audio. The results of EAMM and Wav2Lip under Neutral Base are from \cite{ji2022eamm}. }
    \renewcommand\arraystretch{1.8}
    \tabcolsep=0.05cm
    \begin{tabular}{c|c|c|cccccc|cccccc}
    \hline
    \multirow{2}{*}{\textbf{Method}}&\multirow{2}{*}{\textbf{Avenue}}&\multirow{2}{*}{\textbf{Emotion}}&\multicolumn{6}{c|}{\textbf{Neutral Base}}&\multicolumn{6}{c}{\textbf{Cross Emotion}} \\
    \cline{4-15}
     &&& \textbf{PSNR}$\uparrow$ &\textbf{SSIM}$\uparrow$&\textbf{M-LMD}$\downarrow$ & \textbf{EA$\uparrow$} & \textbf{CSIM}$\uparrow$ &\textbf{LipSync}$\uparrow$&\textbf{PSNR}$\uparrow$&\textbf{SSIM}$\uparrow$&\textbf{M-LMD}$\downarrow$  & \textbf{EA$\uparrow$} & \textbf{CSIM}$\uparrow$&\textbf{LipSync}$\uparrow$ \\
    \hline
    Wav2Lip\cite{prajwal2020lip}&ACMMM'20&\XSolidBrush&29.03&0.57& 3.43 &14.96 & 0.6616&7.36&22.77&0.53&6.93& 11.53& 0.4097&5.74\\
    IP-LAP\cite{zhong2023identity}&CVPR'23&\XSolidBrush&\textbf{32.14}&\textbf{0.97}& 3.52&24.06 & \textbf{0.6790} &\textbf{7.90}&31.06&\textbf{0.97}&4.95 &20.66 &0.4508&5.64\\
    ETK\cite{eskimez2021speech}&T-MM'22&\Checkmark&27.68&0.48&3.73&30.23&0.1234&1.95&10.71&0.55&9.47&10.47&0.1157&1.74\\
    EAMM\cite{ji2022eamm}&SIGGRAPH'22&\Checkmark&29.03&0.66& 2.41& 30.43& 0.2318&5.03&14.52&0.67&8.66& 13.16&0.0349&5.62\\
    SadTalker\cite{zhang2023sadtalker}&CVPR'23&\Checkmark&22.05&0.85&2.71 & 20.66& 0.6080&6.86&22.20&0.73&5.31 &12.44 &0.5686&6.32 \\
    PDFGC\cite{wang2023progressive}&CVPR'23&\Checkmark&20.79&0.69&3.03&27.67&0.4691&6.25&24.85&0.63&4.85&17.73&0.1729&5.80\\
    EAT\cite{gan2023efficient}&ICCV'23&\Checkmark&21.75&0.68&2.45&28.01&0.5052&6.32&23.63&0.67&4.17&23.12&0.5687&6.51\\

    Ground Truth&-&&-&1.00&0.00&60.74&1.0000&-&-&1.00&0.00&60.74&1.0000&- \\
    \textbf{Ours}&-&\Checkmark&31.34&0.96& \textbf{2.39}&\textbf{33.94} & 0.6451&6.94&\textbf{31.63}&\textbf{0.97}& \textbf{2.82}& \textbf{27.49}& \textbf{0.6279}&\textbf{6.76}\\
    \hline
    \end{tabular}

    \label{tab:quantitative}
\end{table*}
\subsection{Datasets.} 
We train our model with LRS2 dataset~\cite{afouras2018deep} and MEAD dataset~\cite{wang2020mead}. LRS2 dataset consists of thousands of spoken sentences from BBC television without emotion annotation. 
This dataset helps our model learn lip-audio synchronization. 
Then, MEAD is introduced to train our cross-emotion memory network. 
The MEAD dataset comprises a collection of talking face videos, showcasing 48 actors and actresses expressing eight distinct emotions. 
From the dataset, we randomly choose 40 actors for training and 8 actors for evaluation. Landmarks are detected for each frame from both datasets with mediapipe tools~\cite{lugaresi2019mediapipe}. The LRS2 dataset is used to train the Audio2Mouth module and Renderer separately.  Then, the MEAD dataset is used to train the Audio Emotion Enhancement module and Emotion Bridging Memory module with the former modules fixed in an end-to-end manner. 

\subsection{Model Structure.} The transformer encoder~\cite{vaswani2017attention} is utilized as our Audio2Mouth module. In it, the feature size is set to 512 and the attention head number is set to 4. The audio encoder is composed of 13 2D convolution layers. The reference encoder and the pose encoder are composed of 1D convolution layers. In the Audio Emotion Enhancement module, the structure of $E_e^a$, $E_t^a$, and $E_c^a$ share the same. They all contain eight 2D convolution layers and three fully connected layers with ReLU activation. For the Emotion Bridging Memory module, the detailed structure of key memory and value memory can be found on External Attention~\cite{guo2022beyond}.

\subsection{Implementation Details.} All videos are cropped to $128\times128$. The sample rate for audio is set to 16kHz. The number of reference images N and the number of adjacent frames predicted at one time T are both set to 5. For Audio2Mouth module, Adam optimizer\cite{kingma2014adam} is utilized with learning rate set to 1e-4. The weight of $L_v$ $\lambda$ is set to 1 during this process. Early Stop is used on L1 loss to decide when to stop training. Finally, the module is trained for 1850 epochs before stopping. For Audio Emotion Enhancement module, following \cite{ji2021audio}, we first pre-train our emotion encoder $E_e^a$ with emotion classification task on MEAD dataset, our timbre encoder $E_t^a$ with identity classification task on MEAD dataset and our content encoder $E_c^a$  with audio2text task on LRS2 dataset. After that, the three pre-trained encoders are trained with the cross-reconstruction strategy\cite{aberman2019learning}, with the learning rate set to $2e-4$. For Renderer, $\omega_{per}$, $\omega_{gan}$ and $\omega_{fm}$ are set to 4, 0.25 and 1 respectively. It takes 7 days to train Renderer with 4 RTX 3090 GPUs. When the modules above are trained, we fix the weights of them to train Emotion Bridging Memory. It takes 60 epochs to train this module.

\subsection{Metrics.} Peak Signal-to-Noise Ratio~(\textbf{PSNR}) and Structured Similarity~(\textbf{SSIM})  are utilized to evaluate the quality of the generated videos. To measure the lip synchronization, we adopt the landmark distance on mouth~(\textbf{M-LMD})~\cite{chen2019hierarchical} and the confidence score of SyncNet~(\textbf{LipSync})~\cite{chung2017out}. Deepface tool~\cite{serengil2021lightface} is applied to assess the emotion accuracy~(\textbf{EA}) of the generated results.  Cosine Similarity~(\textbf{CSIM}) of extracted identity vectors by ArcFace~\cite{deng2019arcface} is calculated to evaluate the identity preservation capability.
\subsection{Comparison with SOTAs}
\begin{figure*}
\includegraphics[width=0.93\linewidth]{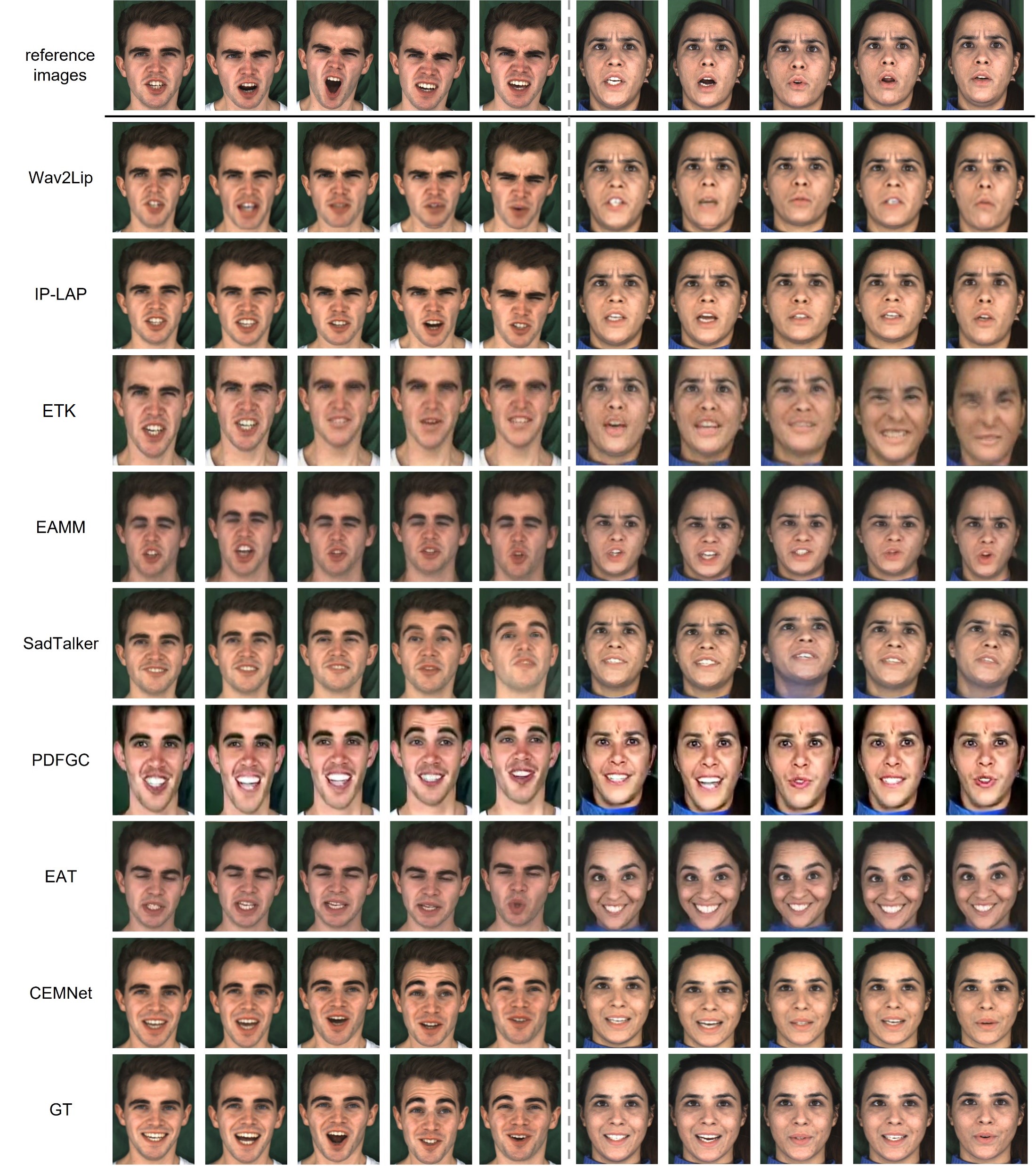}
  \caption{Qualitative comparison with state-of-the-art methods under cross emotion setting on the MEAD dataset. The first row represents the given reference images with 'angry' emotion while the driving audio has 'happy' emotion.}
  \label{figure_qualitative}
  \vspace{-0.7cm}
\end{figure*}
\subsubsection{Comparison Methods.} We perform comparison with state-of-the-art methods on the MEAD and LRS2 dataset. \textbf{Wav2Lip}~\cite{prajwal2020lip} and \textbf{IP-LAP}~\cite{zhong2023identity} are person-generic methods that generate talking face from audio and reference images directly without considering the emotion factor. \textbf{ETK}~\cite{eskimez2021speech} is an emotional talking face animation method which require emotion label as input. \textbf{EAMM}~\cite{ji2022eamm} synthesizes emotional talking face with an additional emotional image as input. \textbf{SadTalker}~\cite{zhang2023sadtalker} realizes emotional face generation by extracting emotion directly from driving audio. \textbf{PDFGC}~\cite{wang2023progressive} presents a progressive disentangled representation learning strategy to achieve fine-grained control over multiple aspects. 
\textbf{EAT}~\cite{gan2023efficient} transforms emotion-agnostic models into emotional ones with parameter-efficient adaptations.

\begin{table}
\centering
\caption{Quantitative comparison with state-of-the-art audio-driven talking face generation methods on LRS2 dataset. The PSNR, SSIM and CSIM metric of Wav2Lip\cite{prajwal2020lip}, IP-LAP\cite{zhong2023identity} and EAMM\cite{ji2022eamm} are from \cite{zhong2023identity}.}
    \renewcommand\arraystretch{1.5}
    \tabcolsep=0.2cm
  \begin{tabular}{c|cccc}
    \toprule
    Method&PSNR$\uparrow$ &SSIM$\uparrow$ & \textbf{SyncScore} $\uparrow$ & \textbf{CSIM} $\uparrow$\\
    \hline
    Wav2Lip\cite{prajwal2020lip}&27.92&0.89&3.86&0.5925\\
    IP-LAP\cite{zhong2023identity}&\textbf{32.91}&\textbf{0.93}&4.49&0.6523\\
    ETK\cite{eskimez2021speech}&13.38&0.5627&2.41&0.2954\\
    EAMM\cite{ji2022eamm}&15.17&0.46&3.01&0.2318\\
    SadTalker\cite{zhang2023sadtalker}&29.48&0.65&5.14&0.5367\\
    PDFGC\cite{wang2023progressive}&23.34&0.64&4.44&0.3217\\
    EAT\cite{gan2023efficient}&16.11&0.65&5.45&0.6180\\
    Ground Truth&-&1.00&8.06&1.0000\\
    \textbf{Ours}&30.19&0.93&\textbf{6.03}&\textbf{0.6609}\\
    \bottomrule

  \end{tabular}
  \label{tab:lrs2_quantitative}
\end{table}
\subsubsection{Quantitative Comparison.} Quantitative comparison with state-of-the-art methods is conducted on MEAD dataset under neutral base setting and cross emotion setting. Only reference images and driving audio from a specific speaker are given in both settings. The reference images are ensured to be neutral emotion in the neutral base setting while in cross emotion setting, reference images are chosen with arbitrary emotion different from that of audio. The ground-truth video has the same emotion and content as the driving audio. We run the evaluation on the test set five times and calculate the average as the final result.

As illustrated in Table~\ref{tab:quantitative}, almost all the methods encounter performance degradation under cross-emotion settings with respect to lip synchronization~(M-LMD, LipSync), identity preservation~(CSIM) and emotion accuracy~(EA). This indicates that the emotional conflicts between reference images and the driving audio will seriously hurt the performance of emotional talking face generation. Among all the methods, our model performs best under the cross-emotion setting. The reason is that our Emotion Bridging Memory can compensate for the lacked motion information and provide accurate landmark displacements from the reference emotion to the target emotion. This illustrates the effectiveness of our proposed memory network CEM-Net.

In neutral base settings, our method achieves the best performance in M-LMD and EA. This illustrates that our lip movement is more consistent with the driving audio and the emotion is more accurate. The cause of this improvement is that even in neutral videos from the MEAD dataset, neutral faces only account for 47.25\% as shown in Fig.~\ref{fig:emo_proportion}, meaning that most neutral faces still have subtle emotion. Methods incorporating target emotion directly into the slightly emotional reference images will encounter emotion accuracy decline and distorted lip movement. 

Quantitative comparison is also conducted on the LRS2 dataset further to evaluate our lip synchronization and identity preservation performance. 
Since there's no emotion annotation for this dataset, we only conduct the common audio-driven talking face generation task. Specifically, for each video, we randomly select 5 frames from it as the reference images and use its audio as the driving audio and the video itself as the ground truth. 
Also, we only calculate the PSNR, SSIM, LipSync, and CSIM to evaluate the lip synchronization and identity preservation ability of our model. The result is presented in Table~\ref{tab:lrs2_quantitative}. As can be seen from the table, our model achieves the best performance on SyncScore and CSIM, which illustrates that CEMNet can generate talking faces with accurate lip movement and high identity preservation.

\begin{figure}[t]
  \centering
\includegraphics[width=\linewidth]{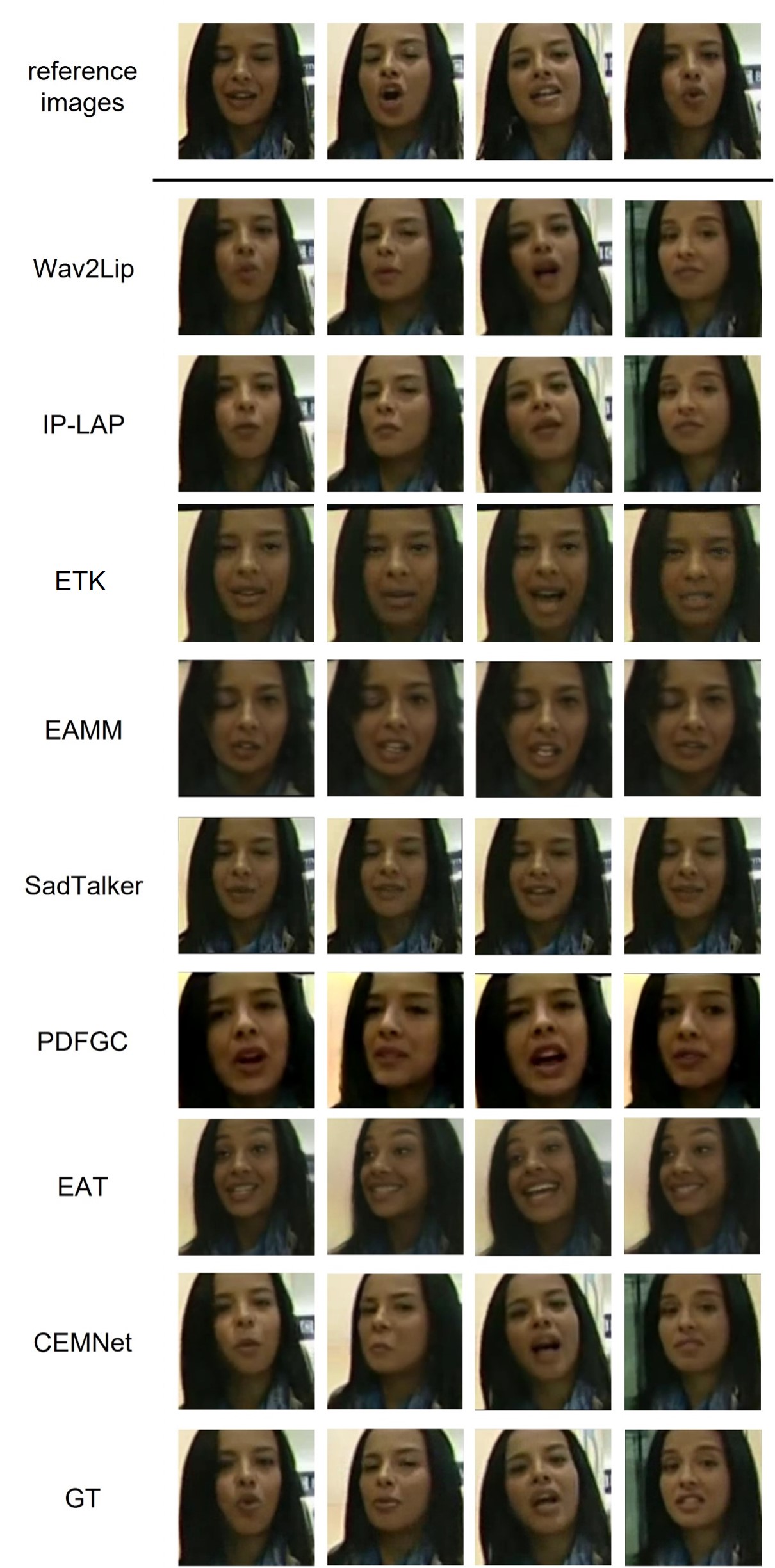}
  \caption{Qualitative comparison with state-of-the-art methods on the LRS2 dataset on the audio-driven talking face generation task. The first row represents the given reference images.} 
  \label{fig:lrs2_qualitative}
\end{figure}
\begin{figure}[t]
  \centering
\includegraphics[width=\linewidth]{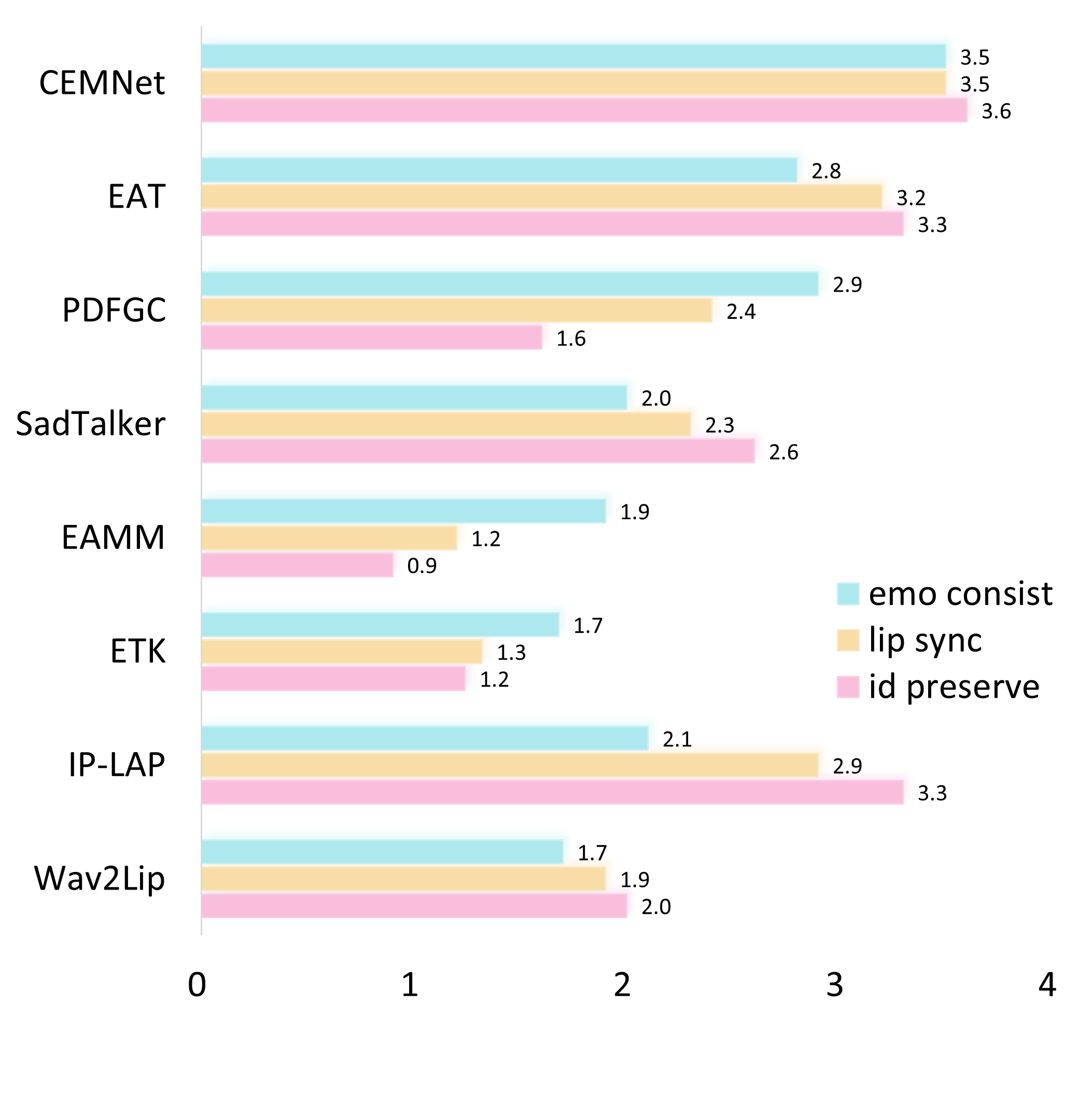}
  \caption{User study results on identity preservation, emotion consistency and lip synchronization.} 
  \label{fig:user_study}
\end{figure}
\subsubsection{Qualitative Comparison.} We first compare our CEMNet with the state-of-the-art methods under the cross-emotion setting on the MEAD dataset. Specifically, two speakers are randomly chosen from the MEAD dataset. We choose the ``angry" emotion as our reference image emotion and the ``happy" emotion as the driving audio emotion. The generated results are represented in Fig.~\ref{figure_qualitative}. 

From the figure, it is visible that the lip movements generated by our model are the closest to GT. Lip movements generated by SadTalker are influenced by the emotion of reference images, mismatching the driving audio. The videos generated by EAMM show adjustments based on the audio emotion ``happy", but since there's no neutral face as input, it directly imposes the ``happy" emotion onto the 'disgusted' reference, resulting in considerable distortion of the mouth shape.
The generated mouth movement of IPLAP is the closest to GT besides our method, but since it does not take emotional factors into account, the lip shapes are still influenced by the emotion from reference images. ETK successfully adds ``happy" emotion to the disgusted face. However, it encounters severe identity distortion. In contrast, our model eliminates emotional information in reference images, making the final generation align well with the driving audio in both emotion and content while preserving the identity features. 

We also choose a speaker from the LRS2 test set and visualize the generated results on the common audio-driven talking face generation task. Specifically, for each video, we randomly select 5 frames from it as the reference images and use its audio as the driving audio and the video itself as the ground truth. The results are shown in Fig.~\ref{fig:lrs2_qualitative}. It can be seen from the figure that our CEMNet can generate the closest lip movement to the GT. However, we noticed that the lips generated by our model are slightly curved downward, displaying a subtle ``disgusted" emotion. This may be because our Emotion Bridging Memory module is trained on the MEAD dataset where clear emotions are presented on the human face. When subtle emotion is detected in the driving audio from the LRS2 dataset, it may exaggerate this subtle emotion in the generated lip shapes. 

\subsubsection{User Study.} It's common practice to validate the effectiveness qualitatively with user study. We conduct a user study following the setting in IP-LAP~\cite{zhong2023identity}. Specifically, 30 individuals participated in the study, all of whom have a basic knowledge of deep learning. For the MEAD test set, one video for each of the eight emotions from the eight speakers is selected. Audios from these videos are extracted to serve as the driving audio. For each driving audio, reference images were randomly obtained from another video of the same speaker with a different emotion. Ultimately, each model generates $8\times8$ emotional videos.  

Participants score the consistency of the generated videos with GT videos from three criteria: lip synchronization, emotion consistency and identity preservation. The scores ranged from 1 to 5, with higher scores indicating better performance. We calculated the average score of the 64 videos for each model as the final score. The result is illustrated in Fig.~\ref{fig:user_study}. Our CEM-Net surpasses other methods in all of the three aspects. It further validates the great capability of our method in cross emotion setting. 

\subsection{Ablation Study}
\begin{figure}[t]
  \centering
\includegraphics[width=\linewidth]{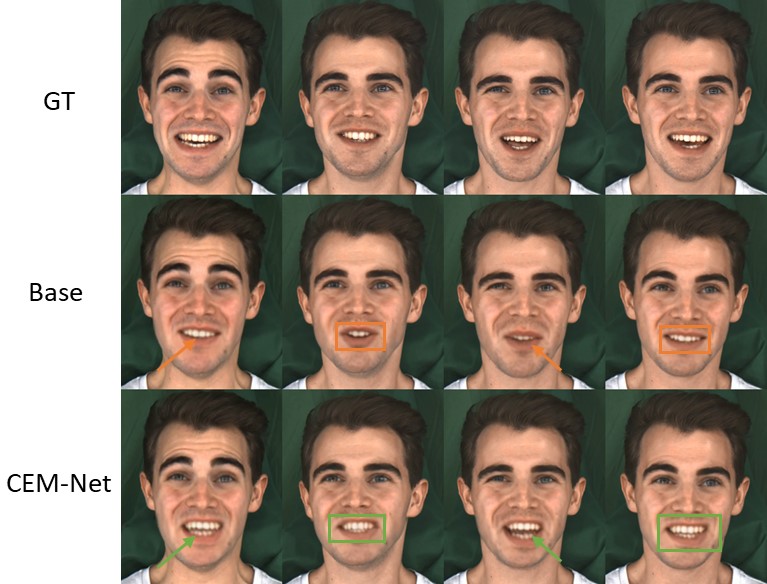}
  \caption{Qualitative results of ablation study. It's clear that our module can help to generate lip movements that match the audio emotion.} 
  \label{fig:ablation_cem}
\end{figure}

\subsubsection{Visualization results.} To validate the effectiveness of our methods to generate emotional lip movements, we visualize the generated results of the baseline and our proposed CEM-Net in Fig.~\ref{fig:ablation_cem}. The baseline is a simple wav2lip model without considering the emotion factor. With the AEE and EBM modules added, our method can generate more expressive and emotional results, especially for the speaker’s mouth corner curvature and lower teeth generation. This indicates the effectiveness of our proposed method.
\begin{table}
\centering
  \caption{Ablation study on two components. We provide quantitative results on M-LMD and EA.}  
\renewcommand\arraystretch{1.5}
    \tabcolsep=0.8cm
  \begin{tabular}{c|c|c}
    \toprule
    Method&M-LMD$\downarrow$ &EA$\uparrow$\\
    \hline
    w/o AEE&3.16&24.38\\
    w/o EBM &4.82&19.53\\
    CEM-Net &\textbf{2.82}&\textbf{27.49}\\
    \bottomrule
  \end{tabular}

  \label{tab:ablation1}
\end{table}

\subsubsection{Audio Emotion Enhancement.} 
To prove necessity of Audio Emotion Enhancement Module~(AEE), we replace our pre-trained audio emotion encoder of AEE with a commonly used emotion classifier\cite{ji2021audio} with the last softmax removed. The comparison result is in Table~\ref{tab:ablation1}. With AEE, there was a significant improvement in the consistency of lip movements and emotion accuracy, which was entirely in line with our expectations. This suggests that our AEE is more accurate in extracting the target emotion compared to the straightforward audio classification task, avoiding the issue of ambiguous emotions caused by person-specific timbre.

\subsubsection{Emotion Bridging Memory.} To verify the effectiveness of our Emotion Bridging Memory(EBM), we attempt to remove the EBM and feed cross-emotion features directly to landmark decoder. This method is indicated as ``w/o EBM". We trained the landmark decoder on the MEAD training set. In Table~\ref{tab:ablation1},  it is observed that removing the EBM witnessed a certain performance degradation in M-LMD and EA. This demonstrates the effectiveness of our EBM in lip-sync and emotion maintenance.

\subsection{Out-Of-Distribution Results.}
\begin{figure}
  \centering
\includegraphics[width=0.9\linewidth]{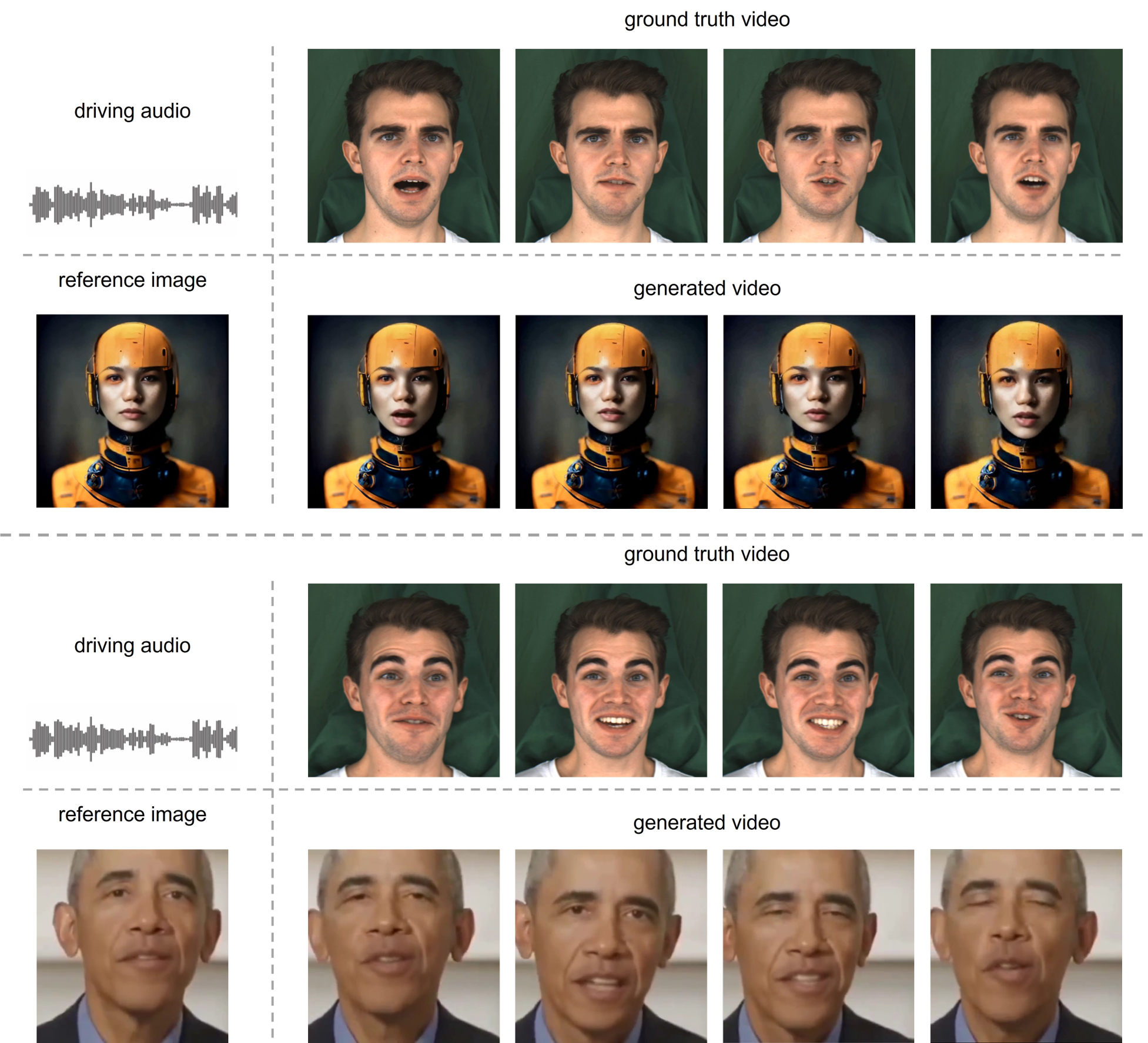}
  \caption{Visualization of out-of-distribution results. We present two out-of-distribution videos, including both cartoon characteristics and real people.} 
  \label{fig:ood}

\end{figure}
In this part, we visualize two examples with out-of-distribution reference images to test the generalization of our proposed method on cartoon characteristics and real people. Audios are chosen from the MEAD dataset with ``neutral" and ``happy" emotions. The reference images are obtained from SadTalker~\cite{zhang2023sadtalker} and Synthesizing Obama~\cite{suwajanakorn2017synthesizing}. The result is shown in Fig.~\ref{fig:ood}. It can be seen from the visualization that the lip movement generated by our model aligns perfectly with the ground-truth videos. This proves that our method generalizes well when it comes to lip synchronization and identity preservation. We also noticed that the lip movements in the generated Obama video exhibit a certain degree of upward curvature. This is because the driving audio contains a 'happy' emotion, and our model successfully incorporated the 'happy' emotion into the generated result.

\section{Conclusion}
In this paper, we present a Cross-Emotion Memory Network(CEM-Net) to realize emotional talking face generation when reference images have strong emotion differing from audio emotion. We employ an Audio Emotion Enhancement~(AEE) module to strengthen the audio emotion. Then an Emotion Bridging Memory~(EBM) module is devised to compensate for the lacked motion information of reference images and speakers' motion habits. Quantitative and qualitative experiments validate that CEM-Net can generate expressive and realistic talking face videos in cross-emotion settings.

\section{Limitations and Future Works.}
While CEM-Net achieves promising results in emotional talking face generation, our method still has limitations. One of the limitations is that we only consider the emotion and lip movements of talking faces to match the driving audio, neglecting other semantic information in audio, such as head pose, which will improve the reality of the generated results further. What's more, due to limited computational resources, we chose to conduct experiments on a baseline that only generates the lower half of the human face to verify the feasibility of the algorithm. However, it's better to generate the entire face in order to fully implement this method.
In future works, we intend to learn head poses from audio to generate natural video. We will also implement our method against baselines that can generate the entire human face to improve our method's usability.

\bibliographystyle{IEEEtran}
\bibliography{main}

\appendix

\section{Experiment Details}
\subsection{Emotion Proportion and Identity Deterioration}
In this part, we present the details in calculating Emotion Proportion for neutral videos in MEAD dataset\cite{ji2022eamm} and LRS2 dataset\cite{son2017lip}, and the evaluation of Identity Deterioration for three face emotion editing methods. 

\textbf{Emotion Proportion}. Deepface\cite{serengil2021lightface} is a lightweight face recognition and facial attribute analysis(age, gender, emotion and race) framework for Python. It can classify a human face into seven different emotions, including angry, fear, neutral, sad, disgust, happy and surprise, with a confidence score for each emotion. To obtain the distribution of each emotion, we first crop and resize each frame from the neutral videos in the MEAD dataset and all videos in LRS2 dataset into $256*256$. After that, we use Deepface to predict the confidence score of seven emotions for each cropped video frame. We calculate the average score of the seven emotions for each dataset. The average score represents the approximate proportion of these emotions in the specific dataset. 

\textbf{Identity Deterioration}.
Arcface\cite{deng2019arcface} is a face recognition network commonly used in image generation tasks\cite{zhong2023identity, kim2022diffusionclip,alaluf2022hyperstyle,park2022synctalkface, roich2022pivotal,deng2022gram,wang2022high} to evaluate the identity preservation ability of generation models. To evaluate the identity preservation capability of face editing models \cite{nitzan2022mystyle, azari2024emostyle, siarohin2019animating}, we first use these models to do face emotion editing with arbitrary images from MEAD dataset. Then, Arcface model is used to extract identity embeddings from the edited image and the original image. Cosine similarity of these two embeddings, marked as $ID$, are then calculated as the identity preservation score. We regard $1-ID$ as Identity Deterioration. So, the closer Identity Deterioration is to 0, the more identity information is preserved.  

\subsection{Implementation Settings.}
In this part, more experimental details are provided. 

\textbf{Model Structure.} We use the transformer encoder\cite{vaswani2017attention} as our Audio2Mouth module. Feature size is set to 512 and attention head number is set to 4. In this part, the audio encoder is composed of 13 2D convolution layers, the reference encoder and the pose encoder is composed of 1D convolution layers. In the Audio Emotion Enhancement module, the structure of $E_e^a$, $E_t^a$ and $E_c^a$ share the same. They all contain eight 2D convolution layers and three fully connected layers with RELU activation. For the Emotion Bridging Memory module, the detailed structure of key memory and value memory can be found on \cite{guo2022beyond}.

\textbf{Training Details.} For Audio2Mouth module, Adam optimizer\cite{kingma2014adam} is utilized with learning rate set to 1e-4. $\lambda$ is set to 1 during this process. Early Stop is used on L1 loss to decide when to stop training. Finally, the module is trained for 1850 epochs before stopping. For Audio Emotion Enhancement module, following \cite{ji2021audio}, we first pre-train our emotion encoder $E_e^a$ with emotion classification task on MEAD dataset, our timbre encoder $E_t^a$ with identity classification task on MEAD dataset and our content encoder $E_c^a$  with audio2text task on LRS2 dataset. After that, the three pre-trained encoders are trained with the cross-reconstruction strategy\cite{aberman2019learning}, with the learning rate set to $2e-4$. For Renderer, $\omega_{per}$, $\omega_{gan}$ and $\omega_{fm}$ are set to 4, 0.25 and 1 respectively. It takes 7 days to train Renderer with 4 RTX 3090 GPUs. When the modules above are trained, we fix the weights of them to train Emotion Bridging Memory. It takes 60 epochs to train this module. 

\section{More results}

\subsection{Comparison Methods Implementation}
We evaluate each method with their publicly available pre-trained models. To test them under the cross-emotion setting, we need to give these methods arbitrary emotion videos or images as reference. And the audio of the ground truth video is given as driving audio. If the testing model is an emotional method, we also give the target emotion of the ground truth video. Specifically, since EAMM\cite{ji2022eamm} is an emotional talking face generation model with an emotional image or emotion label as the target emotion, we set the emotion label input as the ground truth emotion. Because SadTalker\cite{zhang2023sadtalker} can extract target emotion from driving audio, we simply give the ground truth audio as the driving audio. For ETK\cite{eskimez2021speech}, since it can generate videos for seven different emotions simultaneously, we select the generated result with the same emotion as the ground truth video as the final output. Considering IPLAP\cite{zhong2023identity} and Wav2Lip\cite{prajwal2020lip} can only generate the lip movement of the target speaker, we give the upper face of the ground truth video as an additional input. 

\subsection{More cross-emotion results on MEAD dataset}
\begin{figure*}[t]
  \centering
  \includegraphics[width=\linewidth]{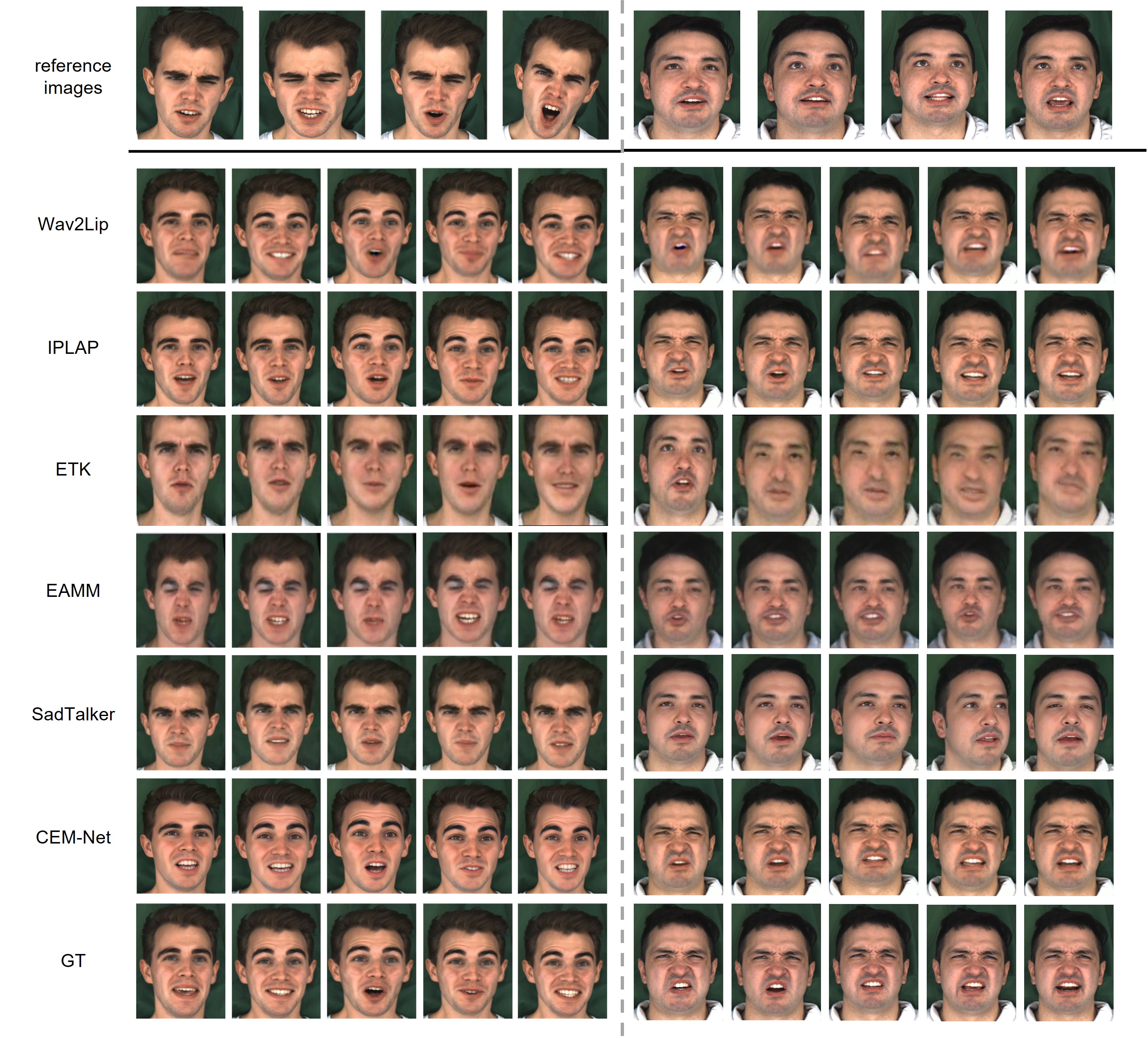}
  \caption{Two speakers from the MEAD testing dataset are selected for qualitative comparison of cross-emotion talking face classification tasks. } 
  \label{fig:sub_qualitative}
\end{figure*}
+
To validate the effectiveness of CEM-Net on solving the cross-emotion talking face generation task, we compare the generated results of CEM-Net on the MEAD dataset\cite{wang2020mead} with State-Of-The-Art methods. Some results are presented in "Qualitative Comparison" section of the paper, and the results of two more speakers are displayed in Figure~\ref{fig:sub_qualitative}. 

In the figure, two speakers are selected from the MEAD testing dataset. For the left speaker, reference images are given with angry emotion, and the driving audio is of happy emotion. For the right speaker, we select reference images from happy emotion videos and the driving audio have disgusted emotion. As can be seen from Figure~\ref{fig:sub_qualitative}, the lip movements generated by our model CEM-Net are the closest to the ground truth video. The results generated by Wav2Lip\cite{prajwal2020lip} are relatively blurry, with considerable distortion in the lip shapes. This is consistent with the lower PSNR and SSIM values we measured in our paper. IPLAP\cite{zhong2023identity} tends to generate lip movements severely influenced by reference images. For example, the generated video for the right speaker is not that disgusted. In fact, the mouth shape of some frames seem to be smiling compared to the ground truth mouth shape. This is illustrated by the lower Emotion Accuracy~(EA) in our quantitative results. The synthesized results of ETK\cite{eskimez2021speech} and EAMM\cite{ji2022eamm} have severely distorted identity information. And the generated emotion is also not correct. This is presented in the low CSIM and EA in our quantitative table. SadTalker\cite{zhang2023sadtalker} can generate quite precise identity information, but the emotion control is not that well. So the EA metric of SadTalker is quite low. Compared to the methods above, our CEM-Net has demonstrated its superiority in cross-emotion talking face generation.

\subsection{Experimantal Results on LRS2 dataset.} 

\begin{table}
\caption{Quantitative comparison with state-of-the-art audio-driven talking face generation methods on LRS2 dataset. The PSNR, SSIM and CSIM metric of Wav2Lip\cite{prajwal2020lip}, IP-LAP\cite{zhong2023identity} and EAMM\cite{ji2022eamm} are from \cite{zhong2023identity}.}
    \renewcommand\arraystretch{1.25}
    \tabcolsep=0.23cm
  \begin{tabular}{c|cccc}
    \toprule
    Method&PSNR$\uparrow$ &SSIM$\uparrow$ & \textbf{SyncScore} $\uparrow$ & \textbf{CSIM} $\uparrow$\\
    \hline
    Wav2Lip\cite{prajwal2020lip}&27.92&0.8962&3.86&0.5925\\
    IP-LAP\cite{zhong2023identity}&\textbf{32.91}&\textbf{0.9399}&4.49&0.6523\\
    ETK\cite{eskimez2021speech}&13.38&0.5627&2.41&0.2954\\
    EAMM\cite{ji2022eamm}&15.17&0.4623&3.01&0.2318\\
    SadTalker\cite{zhang2023sadtalker}&29.48&0.65&4.14&0.5367\\
    Ground Truth&-&1.00&8.06&1.00\\
    \textbf{Ours}&30.19&0.9324&\textbf{5.03}&\textbf{0.6609}\\
    \bottomrule
  \end{tabular}
  \label{tab:sub_quantitative}
\end{table}

To further validate the superiority of CEM-Net on the emotion-agnostic talking face generation task, we test the performance of our model on LRS2 dataset\cite{son2017lip}. 
\begin{figure*}[t]
  \centering
  \includegraphics[width=\linewidth]{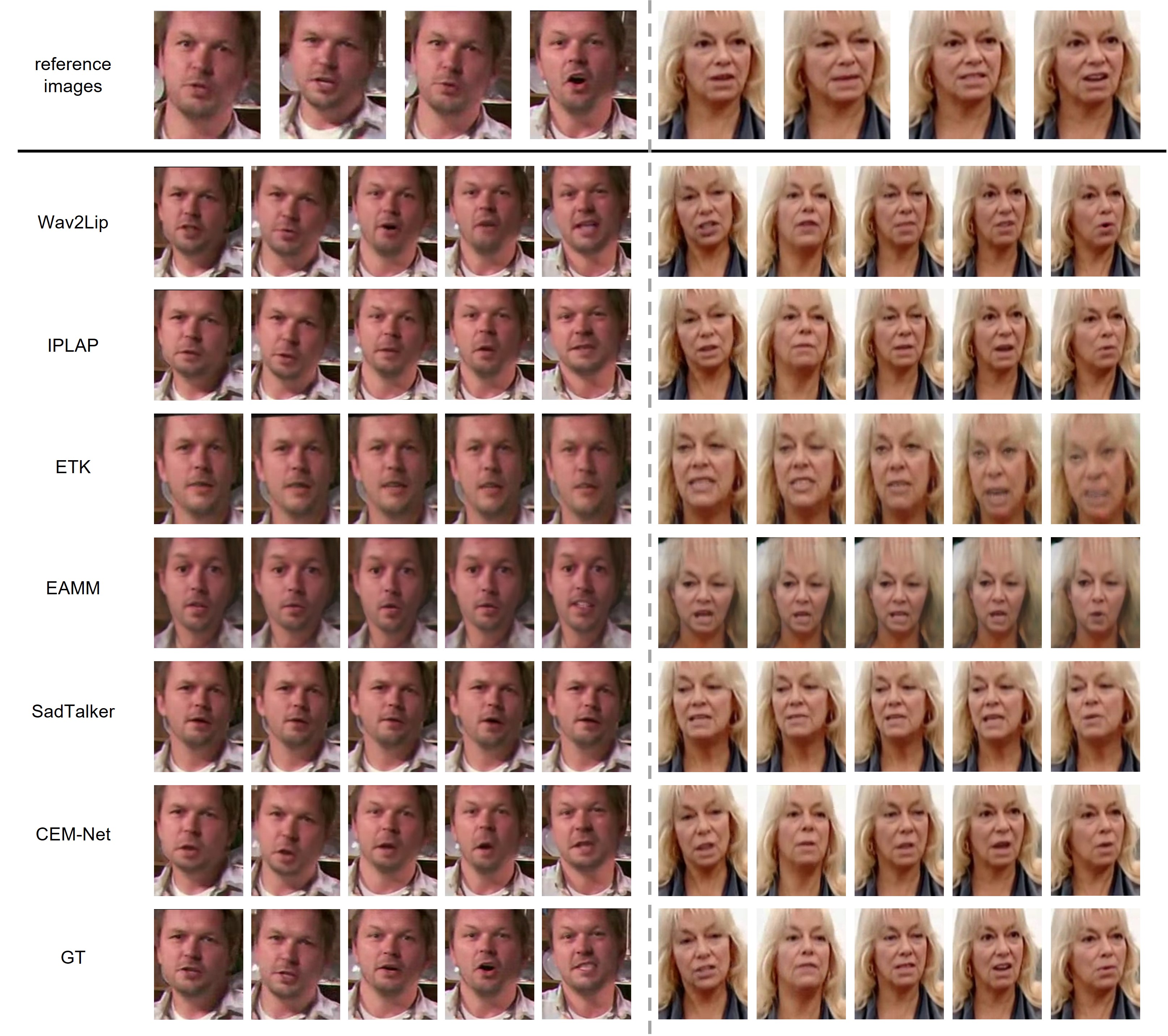}
  \caption{Two video clips from the LRS2 dataset are selected for qualitative comparison of talking face generation task. } 
  \label{fig:lrs_qualitative}
\end{figure*}

\textbf{Metrics.} Peak Signal-to-Noise Ratio~(\textbf{PSNR}) and Structured Similarity~(\textbf{SSIM})  are utilized to evaluate the quality of the generated videos. We also use SyncNet\cite{chung2017out} to calculate SyncScore to measure the synchronization between the generated videos and corresponding driving audios. Cosine Similarity~(\textbf{CSIM}) of extracted identity vectors by ArcFace~\cite{deng2019arcface} is calculated to evaluate the identity preservation capability. 

\begin{figure*}[t]
  \centering
  \includegraphics[width=\linewidth]{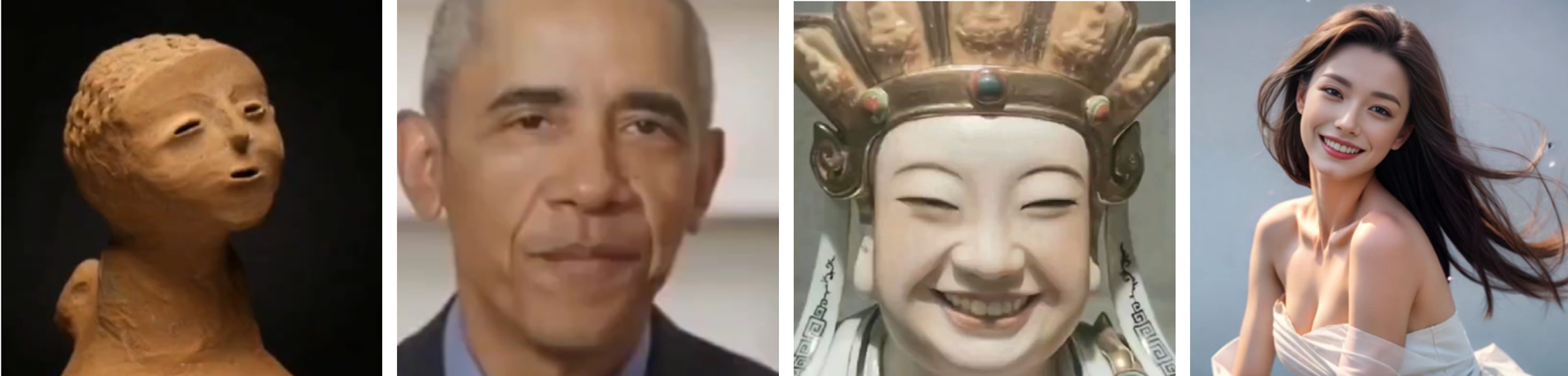}
  \caption{Four in-the-wild images are selected to test the generalization of CEM-Net. The results can be seen in the supplementary video.} 
  \label{fig:in_the_wild}
\end{figure*}
\textbf{Implementation Details.} We randomly selected 45 video clips from LRS2 dataset to evaluate the performance of different methods. Since EAMM\cite{ji2022eamm} requires an driving audio, a reference image, a pose sequence and a driving image as input, we select the first frame of each testing video as the reference image and driving image. The pose sequence of the testing video is extracted as input. For SadTalker\cite{zhang2023sadtalker} and ETK\cite{eskimez2021speech}, we only give the first frame of the video and driving audio as input. We only select the generated video with neutral emotion as the result of ETK. For Wav2Lip, IP-LAP and our method, we give randomly selected images from the testing audio as reference images and the upper face of each frame.

\textbf{Quantitative Comparison.} As can be seen from Table~\ref{tab:sub_quantitative}, our method achieves comparable performance. Since driving audio have the same emotion with reference images in this setting, our CEM-Net achieves only a slight performance improvement in lip synchronization and identity preservation ability with a subtle performance increase. This may be because the randomly selected reference images can have subtle emotion difference with driving audio. And our method can bridge the little emotion gap. Also, since our network can extract additional identity information from reference images and add displacement on the lip landmarks, the mouth movement can be closer to the ground truth speaker.

\textbf{Qualitative Comparison.} 
In Figure~\ref{fig:lrs_qualitative}, we present the visualization of two video clips on LRS2 dataset. From the figure, it is evident that our model has excellent lip generation effects and commendable identity preservation capabilities. This indicates that our model can still produce satisfactory results on emotion-agnostic talking face generation tasks.
\subsection{Experimental Results on Image in the wild}

We also test the generalization ability of CEM-Net on images in the wild. Specifically, four in-the-wild images/videos, as depicted in Figure~\ref{fig:in_the_wild}, are selected as the reference. A piece of audio is randomly chosen from MEAD dataset. The generated results can be seen in the supplementary video. 

\section{Ethical Considerations.}
Audio-driven talking face generation holds great potential for widespread real-world deployments, yet it has the risk of being exploited for generating deepfakes, media manipulation and illicit gains. To counteract such uses, we intend to impose restrictions on the access and use of our code and to embed watermarks in the generated outputs. Furthermore, we are committed to contributing to the deepfake detection community by sharing our synthetic videos, thereby aiding in the enhancement of their detection algorithms. We are convinced that when used ethically, this technology can significantly enrich our daily lives.

\end{document}